\documentclass{ws-procs9x6}

\begin{document}

\title{d+Au Collisions at STAR}

\author{C.A. Gagliardi$^1$,
for the STAR Collaboration\footnote{http://www.star.bnl.gov}}

\address{$^1$\,Cyclotron Institute, Texas A\&M University \\
College Station, TX, 77843, USA\\ 
E-mail: cggroup@comp.tamu.edu}

\maketitle

\abstracts{
STAR has measured forward $\pi^0$ production in p+p and d+Au collisions at $\sqrt{s_{NN}}=200$ GeV.  The p+p yield generally agrees with NLO pQCD calculations.  The d+Au yield is strongly suppressed at $\langle\eta\rangle=4.0$, well below shadowing expectations.  Exploratory measurements of azimuthal correlations between forward $\pi^0$ and mid-rapidity charged hadrons show a recoil peak in p+p that is suppressed in d+Au at low pion energy.  These observations are qualitatively consistent with a saturation picture of the low-$x$ gluon structure of heavy nuclei.  Future measurements to elucidate the dynamics underlying these observations are also described.
}

\section{Current Status}

The BRAHMS Collaboration has reported that negative hadron production in the forward direction is strongly suppressed in d+Au collisions relative to p+p collisions.\cite{BRAHMS_PRL}  The data are well described by calculations that treat the Au nucleus as a color glass condensate.\cite{Khar04}  This may provide evidence at RHIC for the onset of gluon saturation at low-$x$ in the Au nucleus.  However, several other mechanisms have also been proposed to describe the BRAHMS results.  For example, an NLO pQCD calculation\cite{Guze04} concludes that the $\langle x_g \rangle$ sampled by the BRAHMS measurement is $\sim$\,0.02, which casts doubt on the saturation explanation.

To investigate the particle production mechanisms at forward rapidity at RHIC, STAR has measured inclusive $\pi^0$ production in p+p and d+Au collisions at $\sqrt{s_{NN}}=200$ GeV and explored azimuthal correlations between identified $\pi^0$ at $\eta$ = 4 and leading charged hadrons at mid-rapidity.

\begin{figure}[t]
  \begin{minipage}{0.49\textwidth}
    {\epsfxsize=\textwidth\epsfbox{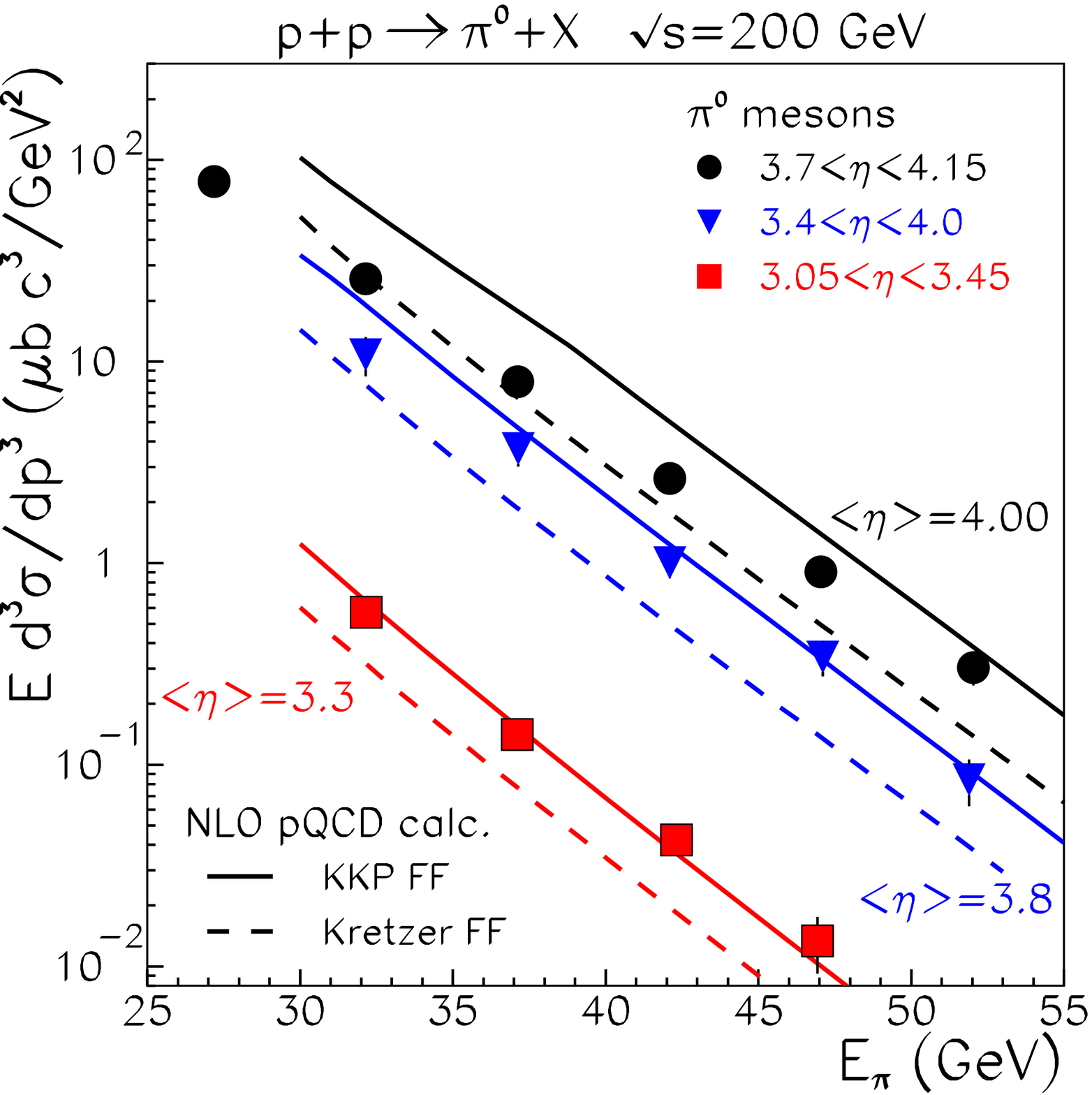}}
  \end{minipage}
  \hfill
  \begin{minipage}{0.49\textwidth}
    {\epsfxsize=\textwidth\epsfbox{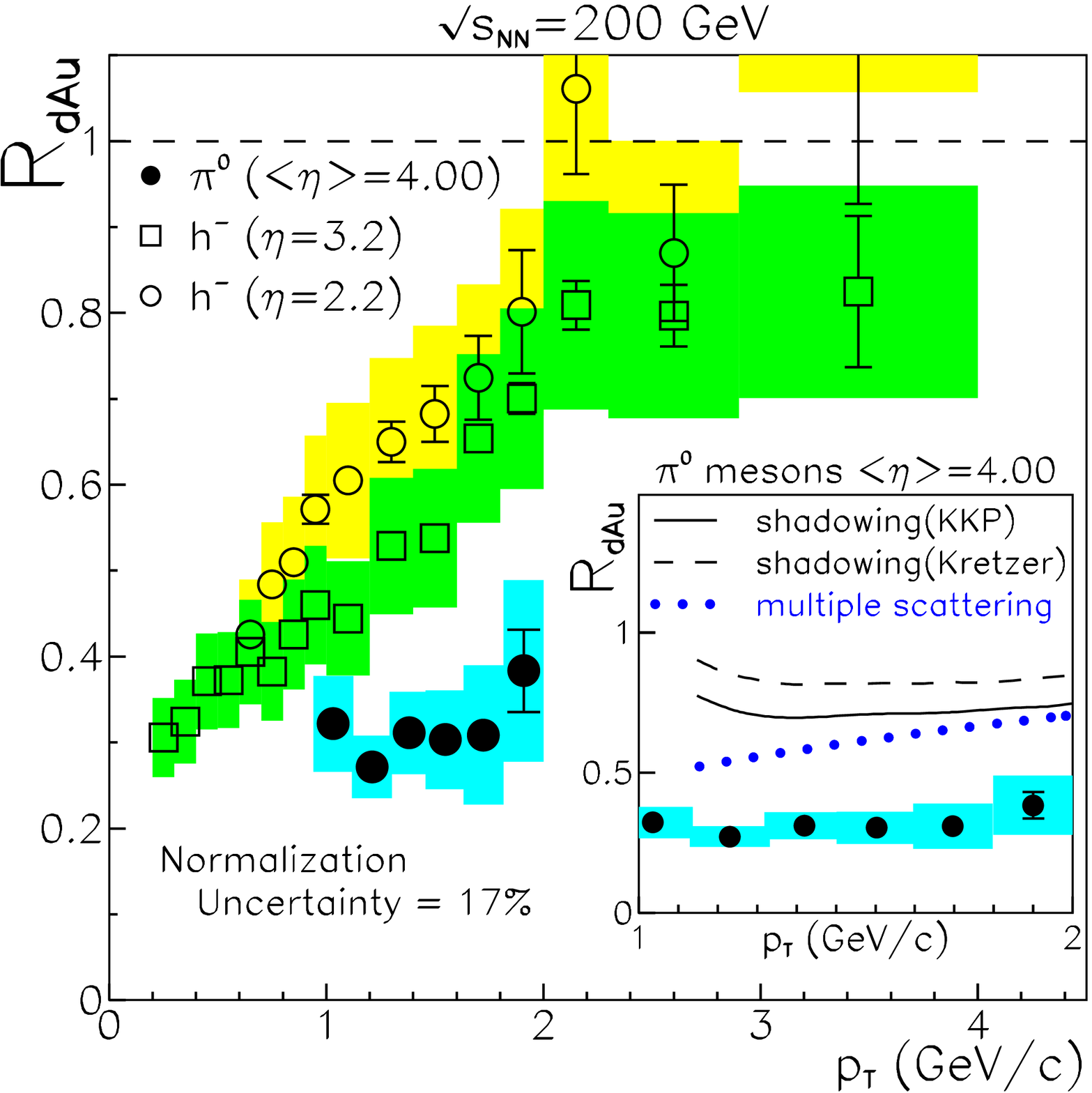}}
  \end{minipage}
\caption{The left panel shows the inclusive $\pi^0$ cross section vs.\@ $E_{\pi}$, compared to NLO pQCD calculations that assume two different fragmentation functions.  Note that $p_T = E_{\pi}/\cosh\eta$.  The right panel shows $R_{dAu}$ for minimum bias collisions vs.\@ $p_T$ for $\pi^0$ at $\langle\eta\rangle=4.0$ from STAR, and for $h^-$ at $\eta=2.2$ and 3.2 from BRAHMS.\protect\cite{BRAHMS_PRL}  The inset compares the measured $\pi^0$ $R_{dAu}$ to various predictions. \label{fig:pi0_yields}}
\end{figure}

Figure \ref{fig:pi0_yields} shows STAR measurements of the inclusive $\pi^0$ production cross sections in p+p collisions at $\sqrt{s}=200$ GeV for $\langle\eta\rangle$ = 3.3, 3.8, and 4.0.\cite{dAu_PRL}  The curves are NLO pQCD calculations using CTEQ6M PDFs and KKP or Kretzer fragmentation functions.  At $\langle\eta\rangle$ = 3.3 and 3.8, the data are consistent with KKP.  At $\langle\eta\rangle$ = 4.0, the data drop below KKP and approach Kretzer as $p_T$ decreases, similar to the behavior observed for mid-rapidity $\pi^0$ production at comparable $p_T$.\cite{PHENIX_pi0}  The NLO pQCD calculations provide a much better description of these data than is found for forward particle production at lower $\sqrt{s}$.\cite{Bour04}

Figure \ref{fig:pi0_yields} also shows the nuclear modification factor for inclusive $\pi^0$ production in d+Au collisions at $\langle\eta\rangle=4.00$, compared to the previous BRAHMS negative hadron measurements at smaller $\eta$.\cite{BRAHMS_PRL}  The STAR $R_{dAu}$ results at $\eta$ = 4 are consistent with a linear extrapolation of the BRAHMS data, if the latter are scaled by 2/3 to account for isospin effects.\cite{Guze04}  The inset demonstrates that the STAR data lie well below shadowing expectations.

\begin{figure}[t]
  \begin{minipage}{0.61\textwidth}
    {\epsfxsize=\textwidth\epsfbox{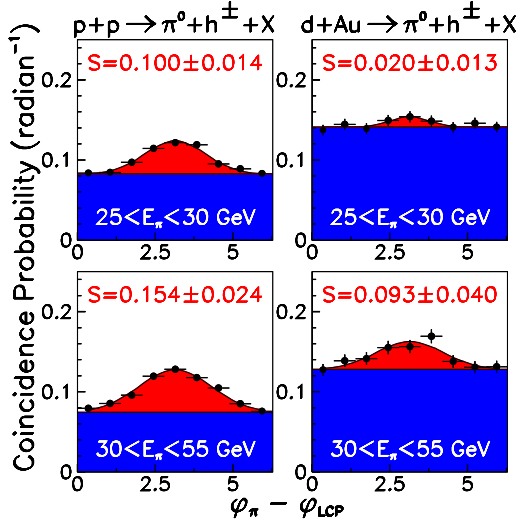}}
  \end{minipage}
  \hfill
  \begin{minipage}{0.34\textwidth}
\caption{Coincidence probability vs.\@ $\Delta\phi$ between a forward $\pi^0$ with $\langle\eta\rangle=4.0$ and a leading charged particle with $p_T>0.5$ GeV/c and $|\eta|<0.75$.  The left (right) column shows p+p (d+Au).  The curves show fits to a Gaussian centered at $\Delta\phi=\pi$ plus a constant background.  $S$ indicates the area of the back-to-back peak.  \label{fig:pi0_h_corr}}
  \end{minipage}
\end{figure}

To explore the dynamics that underlie forward particle production at RHIC, STAR has also studied the azimuthal correlation between forward $\pi^0$ and coincident leading charged hadrons detected at mid-rapidity.  In models that attribute the suppression seen in d+Au collisions to nuclear shadowing or initial-state energy loss, but retain the conventional $2 \to 2$ partonic scattering process, a back-to-back correlation peak should be present for d+Au collisions similar to that found for p+p collisions.  In contrast, in the saturation picture, forward particles arise from energetic quarks in the deuteron that undergo multiple interactions in the dense gluon field of the Au nucleus, leading to an apparent ``mono-jet" mechanism.\cite{Khar03}  Figure \ref{fig:pi0_h_corr} shows the results.\cite{dAu_PRL}  The correlations seen in p+p collisions are very similar to those found in PYTHIA simulations.  HIJING predicts that the back-to-back peak in d+Au collisions should be slightly smaller than that in p+p, with a larger combinatorial background.  In contrast, the back-to-back peak observed in the d+Au data is strongly suppressed at low $E_{\pi}$.  The behavior of the back-to-back peak as a function of $E_{\pi}$ is qualitatively consistent with expectations of the gluon saturation model,\cite{Khar03} but other calculations based on the black-disk limit attribute the results to nearly complete suppression of forward $\pi^0$ production in central d+Au collisions.\cite{Fran06}

\begin{figure}[ht]
  \begin{minipage}{0.52\textwidth}
    {\epsfxsize=\textwidth\epsfbox{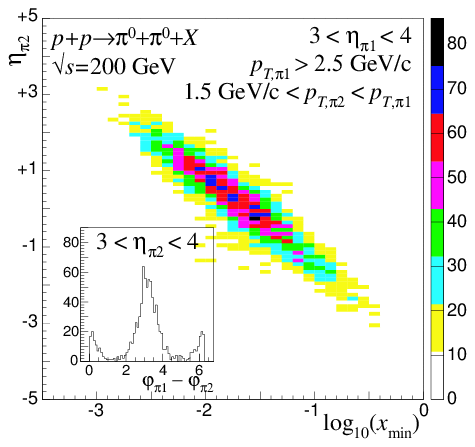}}
  \end{minipage}
  \hfill
  \begin{minipage}{0.46\textwidth}
    {\epsfxsize=\textwidth\epsfbox{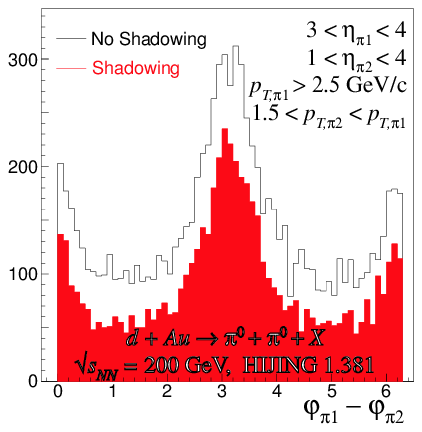}}
  \end{minipage}
\caption{The left panel shows PYTHIA predictions for the correlation between $\eta_2$ and $x_{min}$ for p+p $\to$ $\pi^0 + \pi^0 + X$, where the first $\pi^0$ has $3<\eta_1<4$ and $p_{T,1}>2.5$ GeV/c, and the second $\pi^0$ obeys 1.5 GeV/c $<p_{T,2}<p_{T,1}$.  The inset shows the corresponding $\Delta\phi$ distribution.  The right panel shows similar HIJING predictions for the $\Delta\phi$ distribution in d+Au collisions, with and without shadowing. \label{fig:FMS_sens}}
\end{figure}

\section{Future Plans}

These results are qualitatively consistent with a gluon saturation picture of the Au nucleus, but they cannot rule out other interpretations.  A systematic program of forward measurements is needed to obtain a definitive answer.

STAR is now constructing a Foward Meson Spectrometer (FMS) to enable this program.  The FMS will provide STAR with electromagnetic calorimetry over the range $2.5<\eta<4$, and nearly complete coverage over the full range $-1<\eta<4$ when combined with the existing STAR barrel and endcap electromagnetic calorimeters.  This will permit detailed $\pi^0-\pi^0$ correlations to be observed over a broad kinematic range.  Figure \ref{fig:FMS_sens} shows how back-to-back $\pi^0-\pi^0$ correlations in p+p and d+Au collisions will provide sensitivity to the gluon density in the Au nucleus down to $x_g \sim 10^{-3}$.\cite{Les05}  If the inclusive particle yield in d+Au is suppressed through shadowing, initial-state energy loss, or other mechanisms that retain the conventional $2 \to 2$ partonic scattering process, the back-to-back correlation will be unchanged, as illustrated by HIJING with shadowing.  In contrast, if gluon saturation is the correct explanation, the back-to-back peak will be strongly modified,\cite{Khar03} in addition to the inclusive yield.

\end{document}